\documentclass[11pt,a4paper]{article}
\pdfoutput=1

\usepackage{jheppub}
\usepackage{graphicx}
\usepackage{color}
\usepackage{multirow}
\usepackage{bm}
\usepackage{caption}
\usepackage{subcaption}

\begin{document}

\title{\boldmath Highly excited $B$, $B_s$ and $B_c$ meson spectroscopy from lattice QCD}

\author[a]{Luke Gayer,}
\author[a]{Sin\'ead M. Ryan,}
\author[b]{and David J. Wilson}
\author{\\(for the Hadron Spectrum Collaboration)}

\affiliation[a]{Hamilton Mathematics Institute \& School of Mathematics, Trinity College, Dublin 2, Ireland}
\affiliation[b]{Department of Applied Mathematics and Theoretical Physics, Centre for Mathematical Sciences,\\ University of Cambridge, Wilberforce Road, Cambridge, CB3 0WA, UK}

\emailAdd{lgayer@tcd.ie}
\emailAdd{ryan@maths.tcd.ie}
\emailAdd{d.j.wilson@damtp.cam.ac.uk}

\abstract{
Excited state spectra of $B$, $B_s$ and $B_c$ mesons are computed using lattice QCD. Working with a large basis of carefully constructed, approximately-local $q\bar{q}$-like operators we determine a rich spectrum of states up to $J=4$ in $B, B_s$ and $B_c$, that can be grouped into patterns matching those of the quark model.
In addition, hybrid-like states are identified at approximately 1500 MeV above the ground states in each of $B, B_s$ and $B_c$, in a multiplet pattern already found in similar computations at a range of quark masses spanning light to bottom. This study is performed using an anisotropic, relativistic Wilson-clover action at a single spatial lattice spacing of $a_s\approx 0.12$ fm and with a light-quark mass corresponding to $m_\pi\approx 391$ MeV. We find good agreement with experimental results where known and discuss prospects for further work in interesting $J^P$ quantum numbers.
}

\maketitle

\section{Introduction}
\label{sec:intro}

Spectra of mesons containing a heavy quark are a valuable place to help understand how the quarks and gluons of Quantum Chromodynamics (QCD) bind to form hadrons. The bottom $\bar{b}$ anti-quark when paired with a light, strange or charm quark forms the $B$, $B_s$ and $B_c$ mesons, probing the interactions of QCD at a wide range of energies.

The $B$ and $B_s$ mesons are closely related to the charmed $D$ and charmed-strange $D_s$ mesons, that similarly lie well above the scale of QCD interactions in energy. The presence of light and strange quarks means they interact readily with light mesons through ``connected'' processes and thus strong decay-channel effects are seen everywhere above the ground state pseudoscalar and vector states. The lightest scalar $D_0^\ast$ is thought to have a width of 400 MeV or more~\cite{ParticleDataGroup:2022pth,Du:2020pui,Gayer:2021xzv}. Nevertheless, the $D$ mesons are relatively well studied and many examples have been found experimentally. States such as the scalar $D_0^\ast$ and $D_{s0}^\ast(2317)$ have elucidated how both open and closed channels can affect the observed hadron masses when compared with expectations from quark potential models.

The $B$ and $B_s$ systems might be expected to be similar to the $D$ and $D_s$ with the charm quark replaced by the heavier bottom quark. However, for charmed $B_c$ mesons the picture is very different and may be more like that of charmonium or bottomonium but without charge-conjugation symmetry. Phenomenological quark potential models lead us to anticipate a rich spectrum with possibly several narrow states arising below the lowest ``connected'' $BD$ threshold~\cite{Godfrey:1985xj, Godfrey:2004ya, Lahde:1999ih, DiPierro:2001dwf, Ebert:2009ua, Sun:2014wea, Albaladejo:2016ztm, Lu:2016bbk, Ortega:2016pgg, Ortega:2020uvc, Martin-Gonzalez:2022qwd}. Experimentally, only two $B_c$ states are presently listed by the particle data group (PDG)~\cite{ParticleDataGroup:2022pth} and thus there is much to discover and understand. 

Lattice QCD is a first-principles, systematically improvable approach to QCD whereby correlation functions are computed in a finite, discretised euclidean volume $L^3\times T$. It has proven to be a vital tool in understanding the strong interactions. However, bottom quarks pose a particular challenge since their mass is
typically close to or even above the energy scale defined by the lattice spacing. Several approaches have been developed and utilised in order to make predictions from the fundamental theory. A compendium of previous work using lattice methods includes bottom-(light, strange, charm) spectroscopy with static quarks in Heavy Quark Effective Theory~\cite{Green:2003zza, Koponen:2007nr, Koponen:2007fe, Foley:2007ui, Jansen:2008si, Burch:2008qx, Michael:2010aa, Bernardoni:2015nqa}; Non-Relativistic QCD~\cite{Allison:2004be, Gregory:2009hq, Gregory:2010gm, Wurtz:2015mqa, Mathur:2018epb, Peset:2018jkf}; the Highly Improved Staggered Quark action~\cite{McNeile:2012qf,Lytle:2018ugs}; and the Fermilab approach~\cite{El-Khadra:1996wdx,Oktay:2008ex} in a study of $B_s$ mesons in Ref.~\cite{Lang:2015hza}. 

This work uses a relativistic Wilson-clover action on an anisotropic lattice where the temporal lattice spacing is chosen to be $\sim 3.5$ times finer than the spatial spacing. The same approach has been previously used in an extensive study of bottomonium~\cite{Ryan:2020iog} and is now used in a study of heavy-light mesons, namely the spectra of $B$, $B_s$ and $B_c$ up to $J=4$. Together with relativistic bottom quarks, a large basis of operators enables the identification of several quark model multiplets for each light, strange, and charm flavour.  In addition to these familiar patterns,
multiplets of hybrids are identified, as observed in similar studies using the same methodology for mesons and baryons with other quark flavours~\cite{Dudek:2009qf,Dudek:2010wm,Dudek:2011tt,Edwards:2011jj,Dudek:2011bn,Dudek:2012ag,HadronSpectrum:2012gic,Moir:2013ub,Padmanath:2013zfa,Padmanath:2015jea,Cheung:2016bym,Ryan:2020iog}. We do not include meson-meson-like operators so it is expected that such states will be absent from the spectra. However, they are expected to produce a state when a narrow resonance is present, typically within the width of the resonance~\cite{Dudek:2012xn,Wilson:2015dqa,Gayer:2021xzv}. 

The paper is organised as follows, Section \ref{sec:calc} provides parameters and details of the calculation. Section \ref{sec:fit} gives a description of the fitting and continuum spin identification methods employed. In Section \ref{sec:res} the exotic and excited spectra of $B$, $B_s$ and $B_c$ mesons are presented, and candidate states for a hybrid supermultiplet are identified in each sector. Section \ref{sec:sum} provides a summary and outlook.

\section{Calculation Details}
\label{sec:calc}

In this study, 2+1 flavours of dynamical quarks are used on an anisotropic lattice where the temporal lattice spacing $a_t$ is finer than the spatial lattice spacing $a_s$ and the anisotropy $\xi=a_s/a_t \approx 3.5$. In the gauge sector, a tree-level Symanzik-improved action is used. For the quarks, a tadpole-improved Sheikholeslami-Wohlert (Wilson-clover) action is used, including stout-smeared spatial links ~\cite{Morningstar:2003gk, Sheikholeslami:1985}. Further details of the action and parameter tuning, including the anisotropy and light quark masses, are given in Refs~\cite{Edwards:2008ja,Lin:2008pr}. An approximately-physical strange quark mass is used with a larger-than physical light quark mass, yielding a pion with $m_\pi \approx 391$ MeV when the lattice spacing is set using the $\Omega$ baryon~\cite{Edwards:2012fx}.

While the charm and bottom quarks remain quenched they are simulated with the same Wilson-clover action as the light and strange quarks. The charm-quark tuning is described in Ref.~\cite{HadronSpectrum:2012gic}. The tuning of the bottom quark mass to approximate the experimental $\eta_b$ mass and a mass-dependent determination of the valence anisotropy in the heavy quark sector is described in Ref.~\cite{Ryan:2020iog} and we re-use those parameters here, with two time-sources in a single volume $(L/a_s)^3 \times T/a_t = 20^3 \times 128$. We also briefly consider $(L/a_s)^3 \times T/a_t = 16^3 \times 128$ and $24^3 \times 128$ volumes. Distillation is used to smear the quarks and enable efficient reuse of propagators with different operator constructions~\cite{Peardon:2009gh}.

\subsection{Dispersion relations}
A comparison of lattice dispersion relations for heavy and light mesons provides a useful test of the consistency of the anisotropy tuning and may give indications of the scale of discretisation errors. 
The bottom quark mass and the anisotropy used in this study of heavy-light mesons were tuned in the bottomonium sector, and as discussed in Ref.~\cite{Ryan:2020iog} yielded approximately consistent (relativistic) dispersion relations when used to determine $B$ and $B^\ast$ mesons. Here we present the updated dispersion relations in the vector and pseudoscalar channels for $B, B_s$ and $B_c$ mesons, using two time sources, compared with just a single time source for the heavy-light dispersion relations shown in Ref.~\cite{Ryan:2020iog}.

The relativistic anisotropic dispersion relation for a meson $A$ can be written
\begin{equation}
(a_t E_A)^2 = (a_t m_A)^2 + \frac{1}{\xi_A^2} (a_s p)^2 ,
\label{eqn:Disp}
\end{equation}
where the lattice momenta are quantised so that $a_s\vec{p} = \frac{2\pi}{L} (n_x,n_y,n_z)$ for
$n_i \in \mathbb{Z}$ and are labelled by $[n_x n_y n_z]$ in the following. 

\begin{figure}
\centering
\includegraphics[width = 0.6 \textwidth]{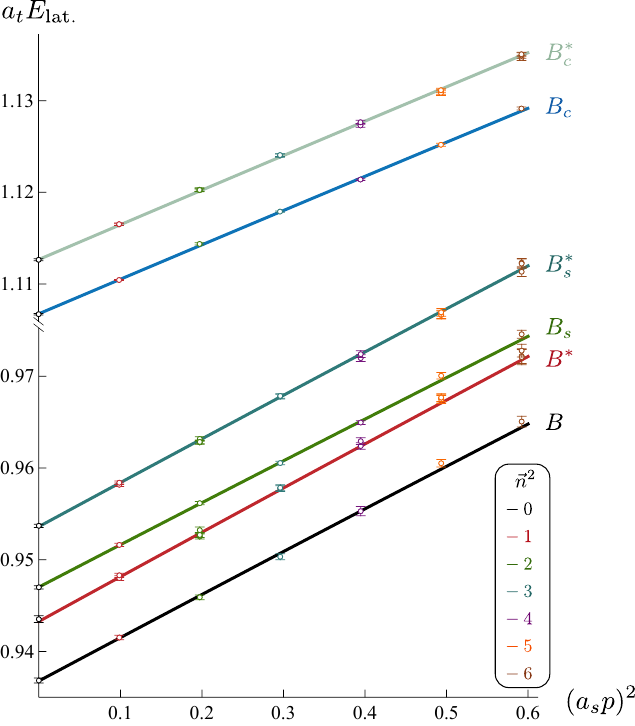}
\caption{Pseudoscalar and vector dispersion relations for $B$, $B_s$ and $B_c$. Momenta up to [211] are shown with both helicity components of the vector states. Lines shown are from fits to Eq.~\ref{eqn:Disp}, including momenta up to [211]. The fit parameters are summarised in Table~\ref{Tab:Disp}.}
\label{fig:Dispersion}
\end{figure}

We determine $m_A$ and $\xi_A$ from fits to Eq.~\ref{eqn:Disp} for both the pseudoscalar and vector $B$, $B_s$ and $B_c$, using discrete values of momentum up to $[211]$, shown in Figure~\ref{fig:Dispersion}. The energies at each momentum are determined from a large basis of interpolating operators using a variational analysis and the fitted anisotropies are given in Table~\ref{Tab:Disp}. The value of the anisotropy in each case is
within $5\%$ of the target value, $\xi_\pi=3.444(6)$ (determined from the pion dispersion relation~\cite{Dudek:2012gj}).
Finally, the same fits were also performed using energy levels with momenta up to $[200]$, and similar results were obtained.\footnote{Small numerical differences with Ref.~\cite{Ryan:2020iog} are due to using two time-sources compared to one in \cite{Ryan:2020iog}.}

Several of the dispersion relation fits have a relatively large $\chi^2/N_\mathrm{dof}$ which is in part due to the small statistical uncertainties on the individual energy levels. Correlations between energy levels are found to be large. For example for the $B_s^\ast$ for $\vec{n}^{\,2}\le 4$, the largest deviation of any level is $1.01\sigma$, while all the others are smaller, yet the $\chi^2/N_\mathrm{dof}$ exceeds 4. This may point to an undetermined systematic uncertainty that is larger than the statistical error. There is also a mild tension between the fitted anisotropies
for different heavy light mesons, while comparing to bottomonium the difference remains small but is statistically significant - as an example
for the  $\eta_b$, $\xi_{\eta_b} = 3.590(15)$~\cite{Ryan:2020iog}. We also compared the measured anisotropies with the same result for the pseudoscalar $B$ meson using up to [200], which yielded $\xi_B=3.397(29)$. The scale of the differences in measured anisotropy may give an indication of the discretisation effects present in the heavy-light sector.

\begin{table}[h]
\begin{center}
\begin{tabular}{c|c|llr}
  &  $\vec{n}^{\;2}\le$ & \multicolumn{1}{c}{$a_tm$}       &  \multicolumn{1}{c}{$\xi$}        &  \multicolumn{1}{c}{$\chi^2/N_\mathrm{dof}$}\\
\hline
$B$         & 4                   & 0.93688(25)  & 3.397(29)    & $\frac{1.33}{5 - 2}  =  0.44$\\ 
            & 6                   & 0.93678(24)  & 3.356(22)    & $\frac{7.24}{7 - 2}  =  1.45$\\
\hline 
$B^\ast$    & 4                   & 0.94332(21)  & 3.306(19)    & $\frac{22.3}{10 - 2}  =  2.78$\\
            & 6                   & 0.94333(21)  & 3.291(18)    & $\frac{30.4}{16 - 2}  =  2.17$\\

\hline 
$B_s$       & 4                   & 0.94704(18)  & 3.388(17)    & $\frac{2.71}{5 - 2}  =  0.90$\\             
            & 6                   & 0.94702(17)  & 3.380(12)    & $\frac{6.59}{7 - 2}  =  1.32$\\
\hline 
$B_s^\ast$  & 4                   & 0.95361(14)  & 3.310(17)    & $\frac{36.0}{10 - 2} = 4.50$\\
            & 6                   & 0.95361(14)  & 3.302(16)    & $\frac{48.5}{16 - 2} = 3.46$\\
\hline 
$B_c$       & 4                   & 1.10675(6)   & 3.461(6)     & $\frac{13.3}{5 - 2}  =  4.43$\\
            & 6                   & 1.10674(6)   & 3.459(6)     & $\frac{21.0}{7 - 2}  =  4.20$\\
\hline 
$B_c^\ast$  & 4                   & 1.11267(8)   & 3.445(12)    & $\frac{17.5}{10 - 2}  =  2.19$\\
            & 6                   & 1.11267(8)   & 3.441(10)    & $\frac{22.0}{16 - 2}  =  1.57$
\end{tabular}
\end{center}
\caption{The parameters determined for each of $B$, $B_s$ and $B_c$ from dispersion relations that include momenta up to [200] ($\vec{n}^{\;2}\le 4$) and up to [211] ($\vec{n}^{\;2} \le 6$).
}
\label{Tab:Disp}
\end{table}

\section{Operator construction, fitting and spin identification}
\label{sec:fit}

The methods utilised in this work have previously been applied to hadrons with light, strange and charm quarks. The notation and methods used are as described in Refs.~\cite{Dudek:2009qf,Dudek:2010wm,HadronSpectrum:2012gic}. Working in a finite cubic volume, rotational symmetry is reduced and we thus work in irreducible representations (irreps) of the cubic group. Meson energies are determined for $B$, $B_s$ and $B_c$ in each lattice irrep by first computing an $N \times N$ matrix of two-point correlation functions,
\begin{equation}
C_{ij}(t) = \langle 0 | \mathcal{O}_i (t) \mathcal{O}_j^{\dagger} (0) | 0 \rangle ,
\end{equation}
where $\mathcal{O}_i$, $\mathcal{O}_j$ are lattice operators in the basis labelled $i,j$, and $N$ is the number of these operators in the irrep, shown in the right side of Table \ref{Tab:irrep}. The operator $\mathcal{O}_{j}^{\dagger}(0)$ creates a state at source time $t=0$, which is annihilated at time $t$ by $\mathcal{O}_{i}(t)$. The correlation function has a spectral decomposition,
\begin{equation}
C_{ij}(t) = \sum_{\mathfrak{n}} \frac{Z_{i}^{\mathfrak{n} *} Z_{j}^{\mathfrak{n}}}{2  E_{\mathfrak{n}}} e^{- E_{\mathfrak{n}} t} ,
\end{equation}
where $Z_i^{\mathfrak{n}} = \langle \mathfrak{n} | \mathcal{O}_i^{\dagger} | 0 \rangle $ are the operator overlaps. 
The lattice operators used in this work are subduced from continuum operators with definite spin and momentum following Ref.~\cite{Dudek:2010wm}. Briefly, the operators are fermion bilinears of the form $\bar{\psi} \Gamma D_i D_j ... \psi $ which are constructed using gamma matrices, $\Gamma$, and forward-backward gauge covariant derivatives, $D_i$, as described in Refs.~\cite{Dudek:2009qf,Dudek:2010wm,HadronSpectrum:2012gic}. A large basis of these operators with up to 3-derivatives is used, including constructions that have previously, and at other quark masses, been shown to overlap well on to states identified as
hybrids. 
The relevant symmetry group for a cubic lattice is $O_h$, with irreps $A_1$, $T_1$, $T_2$, $E$, and $A_2$. Once parity is included, there are ten irreducible representations. The distribution of continuum spin into these irreps, is shown in the left side of Table \ref{Tab:irrep}. 

\begin{table}
\begin{minipage}{0.5\linewidth}
\centering
\begin{tabular}{l| l}
$J$ & $\Lambda$ \\ \hline
0 & $A_1$ \\
1 & $T_1$ \\
2 & $T_2 \oplus E$ \\
3 & $T_1 \oplus T_2 \oplus A_2$ \\
4 & $A_1 \oplus T_1 \oplus T_2 \oplus E$ \\
\end{tabular}
\end{minipage}%
\begin{minipage}{0.5\linewidth}
\centering
\begin{tabular}{l|c c c c c}
$\Lambda$ & $A_1$ & $A_2$ & $E$ & $T_2$ & $T_1$ \\ \hline
$\Lambda^{+}$ & 18 & 10 & 26 & 36 & 44 \\
$\Lambda^{-}$ & 18 & 10 & 26 & 36 & 44 \\
\end{tabular}
\end{minipage}
\caption{The left table shows the distribution of continuum spins up to $J=4$ into the irreducible representations (irreps) of $O_h$, denoted $\Lambda$. The right table shows the number of operators used in each lattice irrep in this study.}
\label{Tab:irrep} 
\end{table}

\begin{figure}
\centering
\includegraphics[width = 0.99 \textwidth]{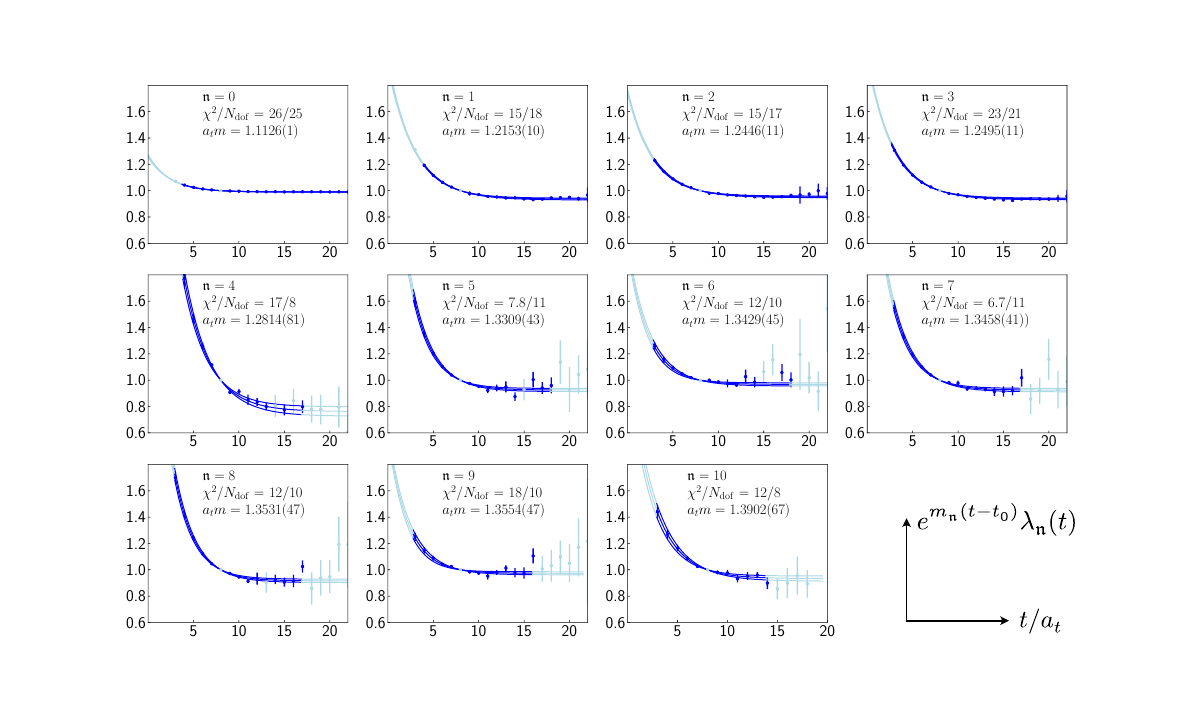}
\caption{Principal correlator fits for the first 11 states, $\mathfrak{n}=0,...,10$, in $T_1^-$ for $B_c$ determined for $t_0/a_t=8$. Darker points are included in the fits while the lighter points with larger statisical
  uncertainties are not included.}
\label{fig:Corr}
\end{figure}

Correlation functions are computed using distillation \cite{Peardon:2009gh}, which enables the use of a large operator basis. By using the variational method~\cite{Michael:1985ne,Luscher:1990ck} energies and operator overlaps can be determined from matrices of correlation functions. In each lattice irrep, a generalised eigenvalue problem (GEVP) is solved,
\begin{equation}
C_{ij}(t) v^{\mathfrak{n}}_j = \lambda_{\mathfrak{n}}(t) C_{ij}(t_0)v^{\mathfrak{n}}_j,
\end{equation}
where $i$ and $j$ label operators and $t_0$ is the reference timeslice. Solving the GEVP for each $t$ gives eigenvalues, $\lambda_{\mathfrak{n}}$ and eigenvectors $v^{\mathfrak{n}}_j$. The eigenvectors are related to the operator overlaps $Z_i^{\mathfrak{n}}$, which enable the identification of the continuum spin of a state, as described below. The eigenvalues, or principal correlators, $\lambda_{\mathfrak{n}}$ are fitted to the form
\begin{equation}
\lambda_{\mathfrak{n}}(t) = (1-A_{\mathfrak{n}}) e^{-m_{\mathfrak{n}}(t-t_0)} + A_{\mathfrak{n}} e^{-m'_{\mathfrak{n}} (t-t_0)} ,
\end{equation}
with free parameters $m_{\mathfrak{n}}$, $m'_{\mathfrak{n}}$ and $A_{\mathfrak{n}}$. Here $m_{\mathfrak{n}}$ is the energy of the state of interest, and the other free parameters help account for contamination by excited states, and are not relevant to the subsequent analysis. An example of the fits performed, in this case for the first 11 states of $T_1^-$ in $B_c$, is shown in Figure \ref{fig:Corr}, which shows $e^{m_{\mathfrak{n}}(t-t_0)} \lambda_{\mathfrak{n}}(t)$ plotted against $t/a_t$. In general a good signal is observed that is stable against variation of $t_0$ and with reasonable $\chi^2/N_{\mathrm{dof}}$. Similar quality fits are obtained in $B$ and $B_s$, and across all the irreps.

\subsection{Spin identification}
\label{sec:spinid}
Lattice energy levels are assigned a continuum spin following the approach described in Refs.~\cite{Dudek:2009qf,Dudek:2010wm}. For states with continuum spins which subduce into a single irrep, $J = (0,1)$, the operator overlaps, $Z_i^{\mathfrak{n}}$, alone can indicate the assignment. More generally, by comparing the overlap values of energy levels in different lattice irreps which have been subduced from the same continuum operator, a continuum spin assignment can be made~\cite{Davoudi:2012ya}. This method has proven to be a incisive tool for spin identification, especially for excited states and in cases where a dense spectrum of energy levels means that discriminating between fitted energies alone is not reliable.

\begin{figure}[b]
\centering
\includegraphics[width=0.99 \textwidth]{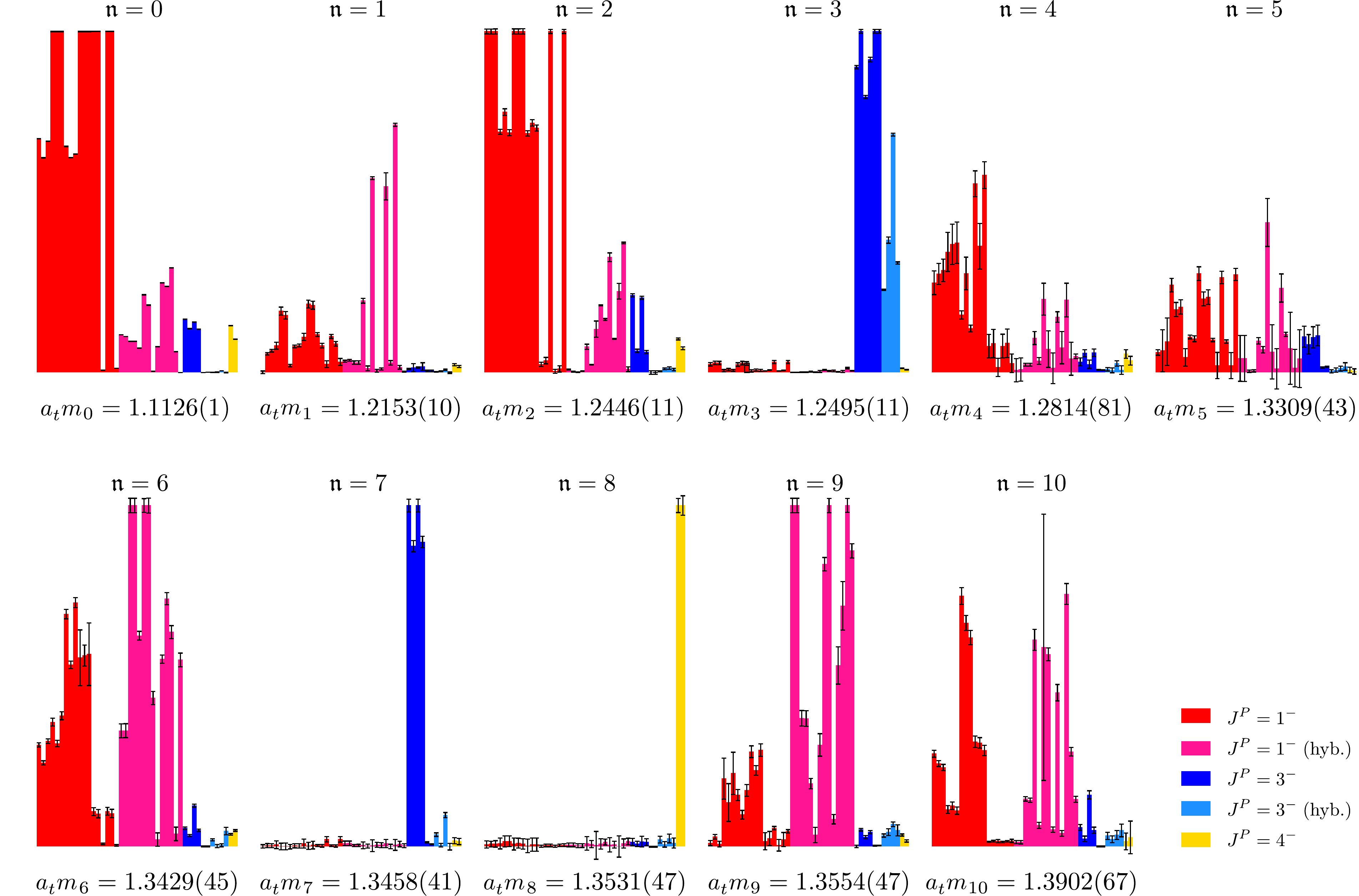}
\caption{The normalised operator overlaps $|\tilde{Z}|$ of the first 11 states, $\mathfrak{n}=0,...,10$, determined in the $T_1^-$ irrep of the $B_c$ meson. Operator overlaps are normalised relative to the largest overlap among all states. The operators are coloured according to their continuum spin, with lighter shades of each colour representing hybrid-like operators. The mass, $a_t m$, of each state determined in principal correlator fits is also shown.}
\label{fig:Zhist}
\end{figure}

The operator overlaps of the first 11 states in the $T_1^-$ irrep of $B_c$ are shown in Figure \ref{fig:Zhist}. The colour coding highlights the continuum spin from which
each operator in a level was subduced and the lighter shades of each colour denotes operators that are expected to
overlap strongly onto hybrid states.

\begin{figure}[t]
\centering
\includegraphics[width = 0.75 \textwidth]{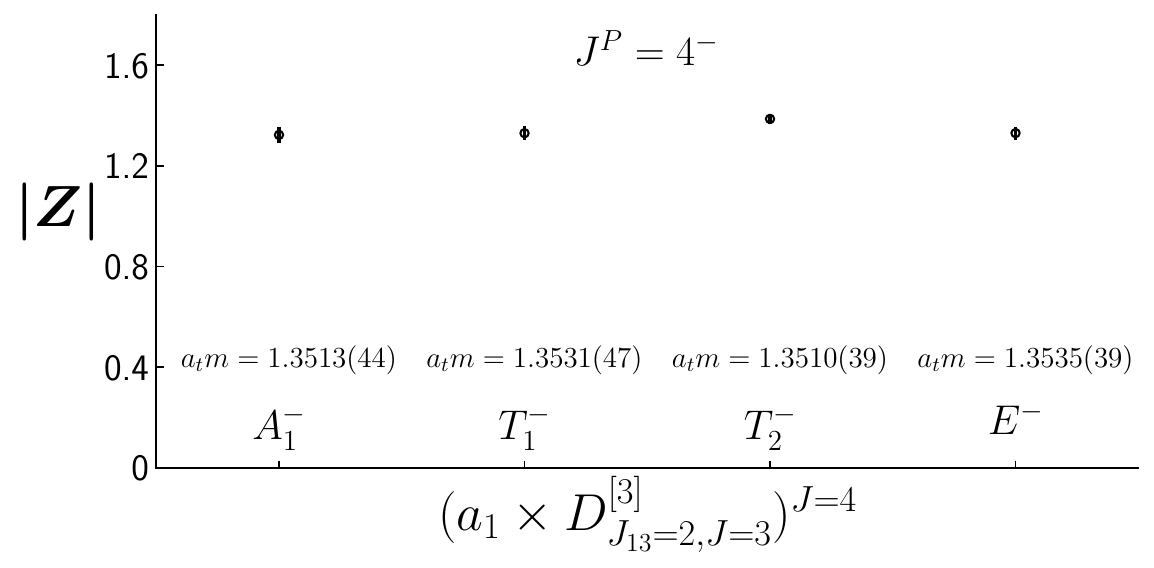}
\caption{Operator overlaps for an operator with continuum $J^P = 4^-$ subduced into the $A_1^-$, $T_1^-$, $T_2^-$ and $E^-$ lattice irreps in $B_c$. The lattice operator naming convention follows Ref.~\cite{Dudek:2010wm}. The masses and overlaps are consistent across the four irreps. These four lattice states are therefore identified as the same continuum $4^-$ state.}
\label{fig:SpinId}
\end{figure}

In Figure \ref{fig:SpinId} the identification of a highly-excited state using the combination of operator overlaps and the fitted energy levels is illustrated. The overlaps of an operator with $J^P = 4^-$ subduced into each of the lattice irreps $A_1^-$, $T_1^-$, $T_2^-$ and $E^-$ are shown. Since the overlaps and masses obtained are consistent across the four irreps, these lattice states are identified as the same continuum state with $J^P = 4^-$. The same process is used to identify all continuum states shown in this work with spin $J \geq 2$.

\section{Results}
\label{sec:res}
In this section the results for the spectra of heavy-light mesons, $B,B_s$ and $B_c$, including candidate hybrids, are presented. Results organised by the relevant lattice irreps are shown, followed by the spectra labelled by continuum quantum numbers, and an analysis and comparison of mixing angles.
\subsection{Spectra by Lattice Irrep}
\begin{figure}
	\centering
	\begin{subfigure}[b]{0.99\textwidth}
		\includegraphics{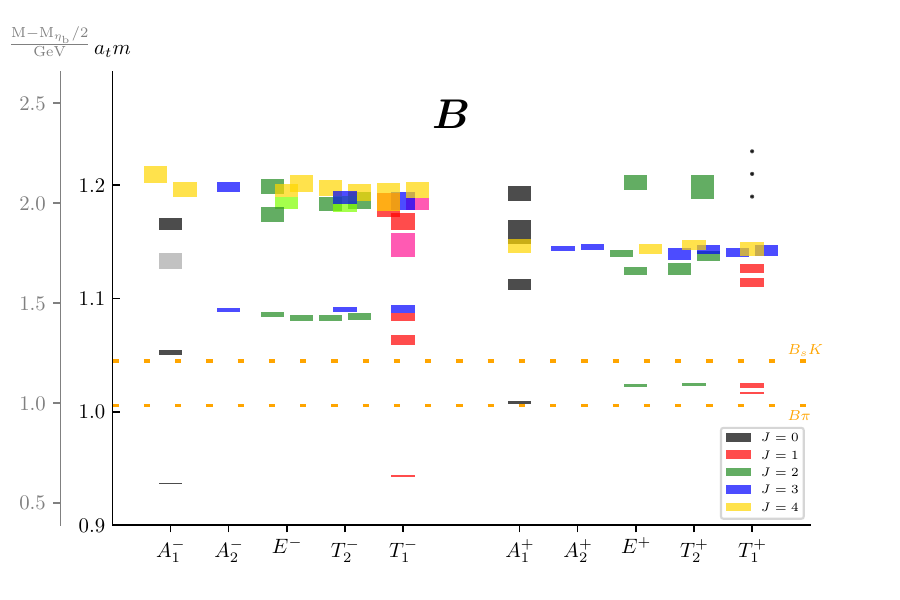}
	\end{subfigure}
        \caption{The spectrum arranged by lattice irrep for $B$. States are colour coded according to the continuum spin of the dominant operator in each level. The vertical height of the boxes indicates the one sigma statistical uncertainty about the mean. Lighter colours in each shade denote a dominant
          hybrid-like operator overlap. Ellipsis indicate that more states may be present at higher energies but were not robustly determined in this study. The orange dashed lines are the lowest OZI connected and OZI disconnected lattice thresholds. }
\label{fig:IrrepSpectra}
\end{figure}

\begin{figure}
	\begin{subfigure}[b]{0.99\textwidth}
		\includegraphics{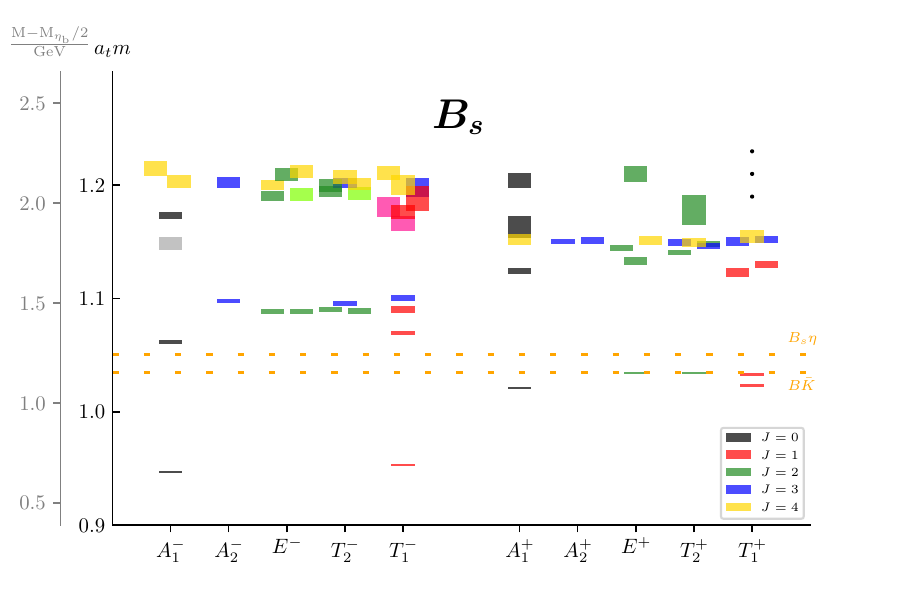}
	\end{subfigure}
	\begin{subfigure}[b]{0.99\textwidth}
		\includegraphics{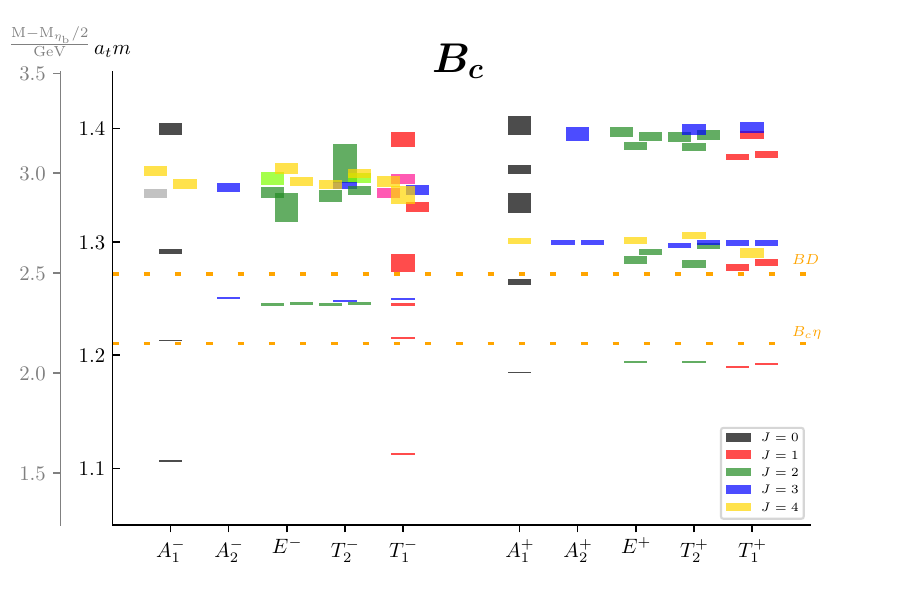}
	\end{subfigure}
\caption{As for Figure \ref{fig:IrrepSpectra} but for $B_s$ (upper plot) and $B_c$ (lower plot).}
\label{fig:IrrepSpectraBc}
\end{figure}

The spectra of excited and exotic $B$, $B_s$ and $B_c$ mesons arranged by lattice irrep is shown in Figures~\ref{fig:IrrepSpectra} and~\ref{fig:IrrepSpectraBc}. Results are presented as mass splittings, with half the mass of the $\eta_b$ meson subtracted to mitigate uncertainties arising from the tuning of the bottom quark. The energy levels shown are colour-coded according to the continuum spin of the dominant operator overlap at each level, with black for $J=0$, red for $J=1$, green for $J=2$, blue for $J=3$ and gold for $J=4$, using the method described in Section~\ref{sec:spinid}. The pattern of states observed is similar across the three flavour sectors.

\begin{figure}[t]
\centering
\includegraphics[width=0.99\textwidth]{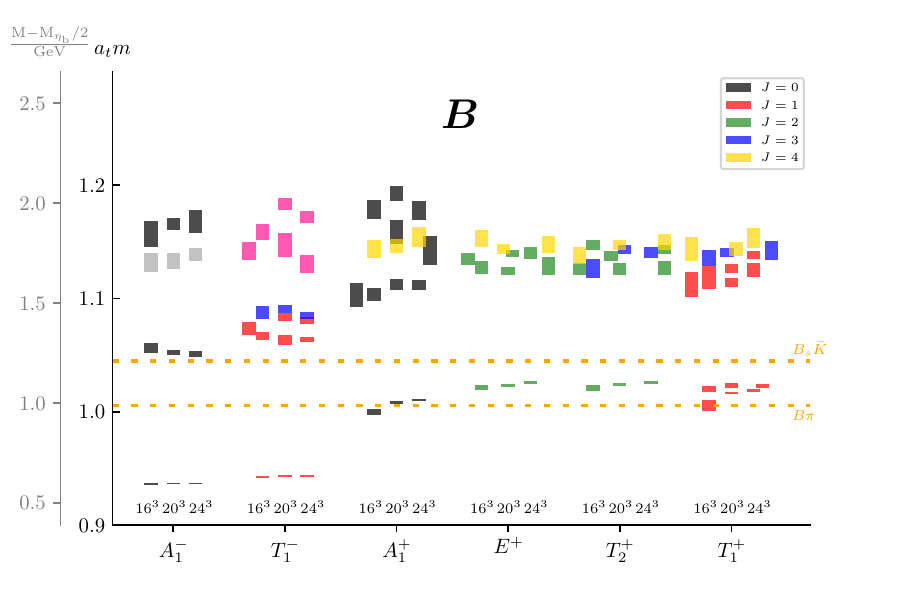}
\caption{As for Fig.~\ref{fig:IrrepSpectra}, for a selection of irreps in the B meson spectrum at three volumes, as described in the text.}
\label{fig:Vol}
\end{figure}

To investigate the volume dependence of these results the $B$ meson spectrum was computed at two additional volumes. In Figure \ref{fig:Vol}, a comparison of the $B$ meson spectrum at three volumes $L/a_s = 16, 20, 24$, is presented. The irreps which contain the lowest five states are shown. The overall pattern and ordering of states is consistent across the three volumes. Nevertheless, while the lightest states in $A_{1}^{-}$, $T_{1}^{-}$, $E^+$ and $T_{2}^{+}$ are in agreement a slight shift with volume is observed in the lightest $A_{1}^{+}$ and $T_{1}^{+}$ states, corresponding to the lightest $J^P = 0^+ , 1^+$ states. A similar effect was found in the $D$ meson sector where finite volume effects in these channels were also relatively large and
subsequent scattering analyses including meson-meson-like operators in $D \pi$ and $D^\ast \pi$ channels~\cite{Moir:2016srx,Gayer:2021xzv,Lang:2022elg}
identified a near-threshold bound-state strongly-coupled to the meson-meson channel.

\subsection{The Spin-Identified Continuum Spectra}

\begin{figure}
	\centering
	\begin{subfigure}[b]{0.99\textwidth}
		\includegraphics{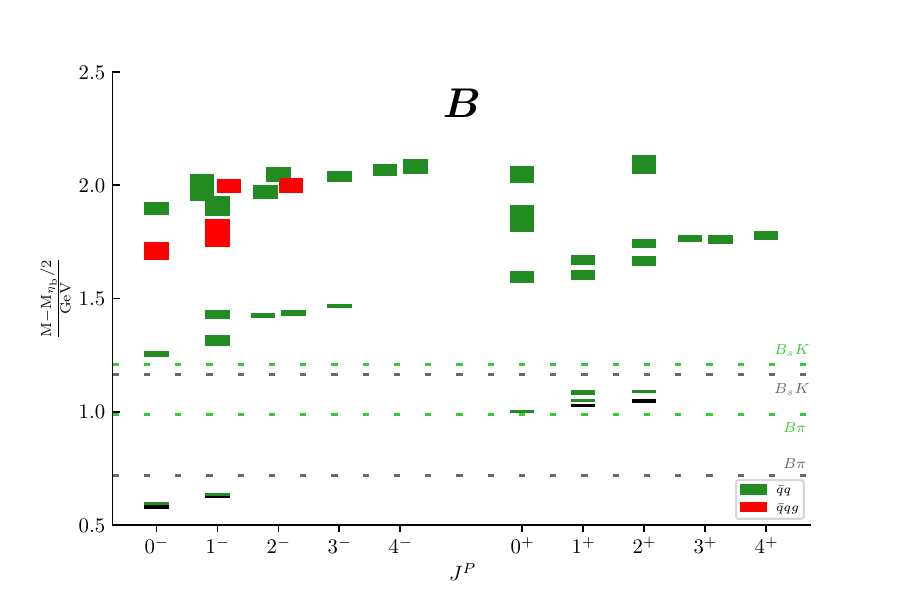}
	\end{subfigure}
        \caption{Spin-identified spectra labelled by continuum $J^P$ for $B$ mesons. The vertical height of the boxes represents the one sigma statistical uncertainty. States in red and green have been determined in this study, while states in black are experimental measurements. The states shaded
          red are dominated by operators with a hybrid-like construction. Dashed lines show the lowest relevant decay thresholds, green determined in this work while black indicates the experimental thresholds.}
        \label{fig:ContSpectra}
\end{figure}

\begin{figure}
  \begin{subfigure}[b]{0.99\textwidth}
    \includegraphics[width=0.99 \textwidth]{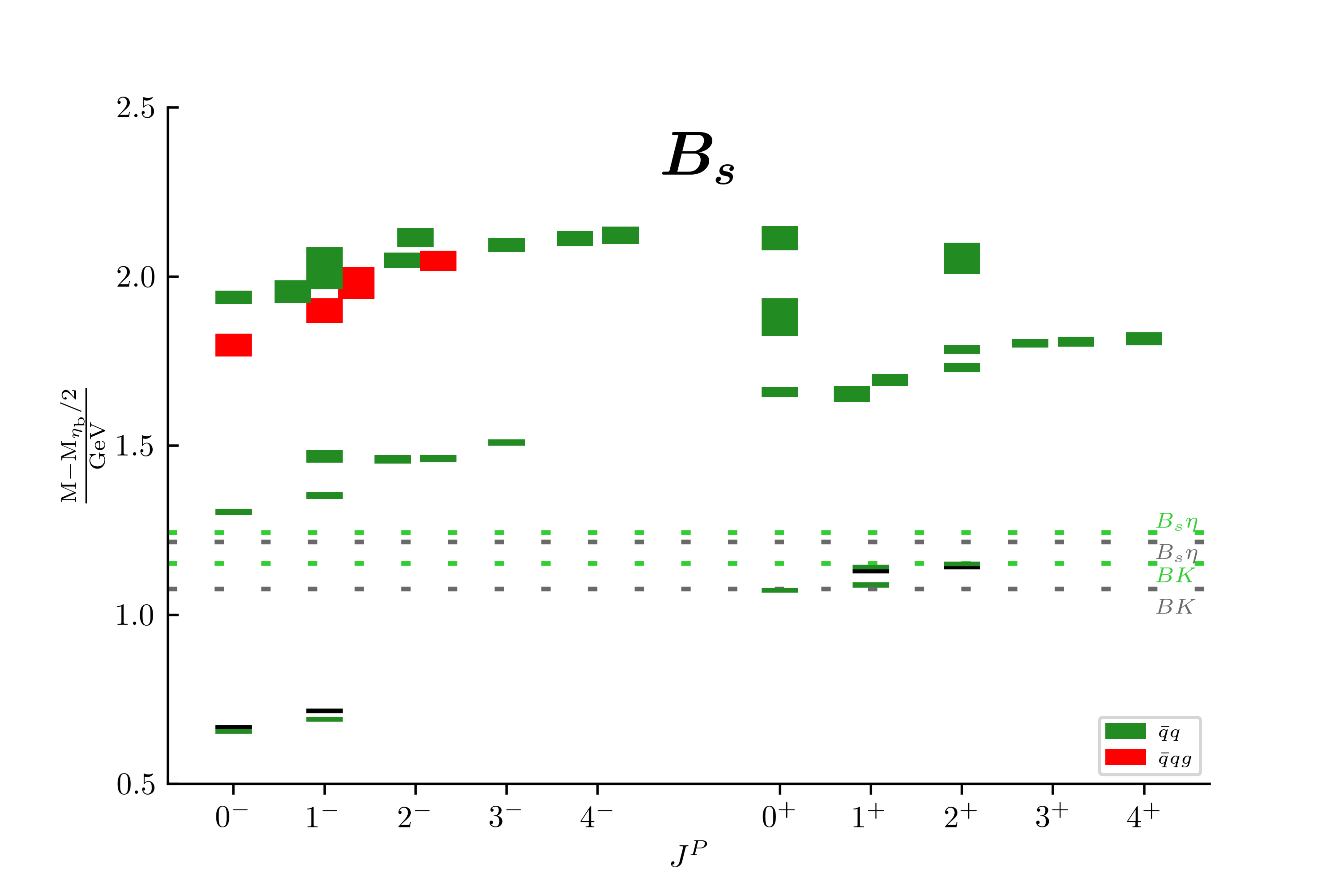}
  \end{subfigure}
  \begin{subfigure}[b]{0.99\textwidth}
    \includegraphics{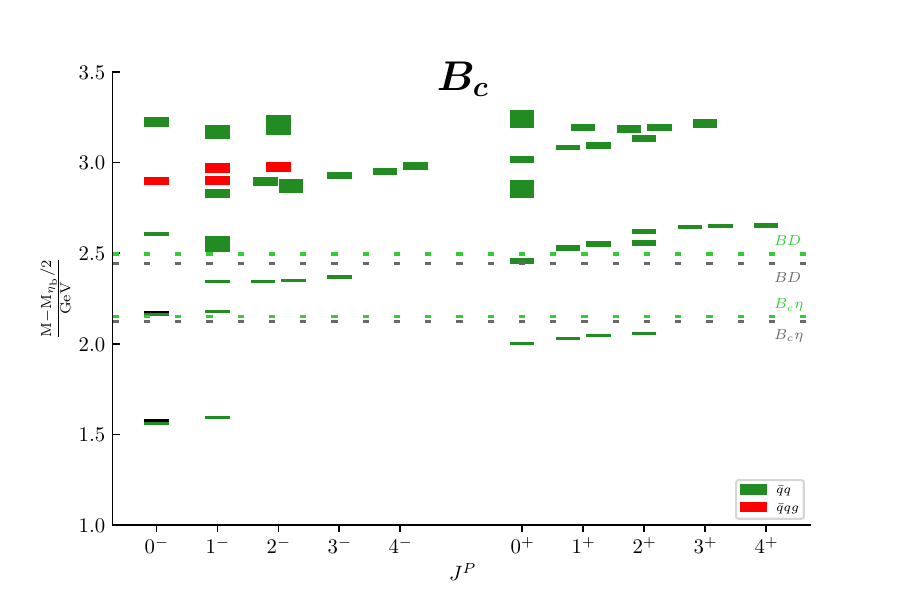}
  \end{subfigure}
  \caption{As for Figure~\ref{fig:ContSpectra} but for the $B_s$ (upper plot) and $B_c$ (lower plot).}
  \label{fig:ContSpectra2}
\end{figure}

The spin-identified spectra labelled by continuum $J^P$ are shown in
Figures~\ref{fig:ContSpectra} and~\ref{fig:ContSpectra2}. Spin identification has been done as described in Section~\ref{sec:spinid} and masses are presented with half the mass of the $\eta_b$ meson subtracted. The pattern of states is similar in the
different flavour sectors and is
similar to patterns of states determined in bottomonium \cite{Ryan:2020iog}, charmonium \cite{Moir:2013ub}, and charmed mesons \cite{HadronSpectrum:2012gic}.

\begin{figure}[tb]
\centering
\includegraphics[width=0.99\textwidth]{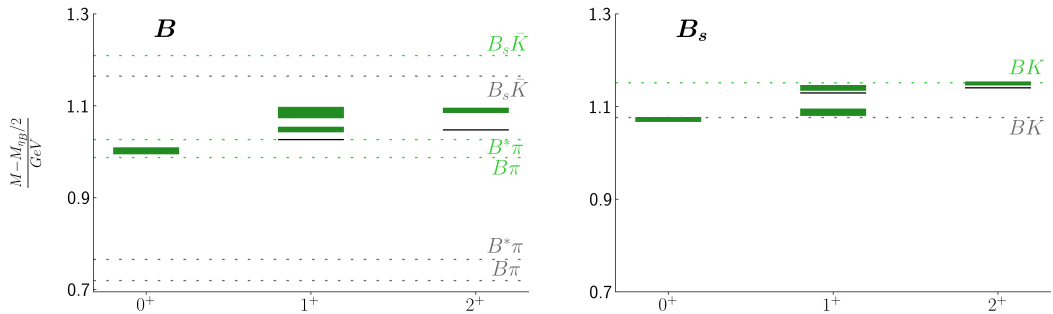}
\caption{A detailed view of the low lying positive parity states in $B$ and $B_s$. The energy levels and thresholds determined here are shown in green, while experimental states and thresholds are marked in black. The proximity of these states to decay thresholds makes them particularly interesting and worthy of further study.}
\label{fig:Zoom}
\end{figure}

In $B$ and $B_s$, there are a number of states with positive parity which are near relevant thresholds. Figure \ref{fig:Zoom} highlights the detail
of these near-threshold states with $J^P = (0,1,2)^+$.\footnote{One expects a meson-meson state to play a role around these energies and so a more complete study including relevant meson-meson operators will be essential for a full understanding of the interactions in these quantum numbers.}
Looking more closely at the $B$ mesons, the lowest-lying state with $J^P = 0^+$ is found $15 \pm 8$ MeV above the lattice $B \pi$ threshold, and strong meson-meson effects are expected. The ground state in $J^P = 1^+$ is $22 \pm 6$ MeV above the $B^* \pi$ threshold. In $B_s$, the second state with $J^P = 1^+$ is approximately $11 \pm 6$ MeV below the lattice $BK$ threshold, while the first state with $J^P=2^+$ lies within $1\pm 4$ MeV of
the lattice $BK$ threshold.
A comparison to the $D$~\cite{Moir:2016srx,Gayer:2021xzv,Lang:2022elg} and $D_s$~\cite{Cheung:2020mql} systems is instructive given the similarities to Ref.~\cite{Moir:2013ub}, and we might therefore anticipate bound-states in both $B$ and $B_s$ for $0^+$ and $1^+$ once meson-meson operators are included, bearing in mind that both studies use the same unphysically heavy pion, $m_\pi\sim 391$ MeV. 

In Table \ref{Tab:Splittings}, a selection of splittings across the three meson sectors is presented and compared to experimental values from the PDG~\cite{ParticleDataGroup:2022pth}. We note that while the hyperfine splittings are generally underestimated, a similar effect was found in Refs.~\cite{HadronSpectrum:2012gic,Moir:2013ub,Cheung:2016bym} using the same heavy-quark action.
The value of the spatial clover coefficient, $c_s$ in the fermion action is known to effect the value of hyperfine splittings in a lattice calculation. In this study, $c_s$ takes its tree-level tadpole improved value. Increasing $c_s$ to approximate its non-perturbative value has been shown to improve the determination of the hyperfine splitting in charmonium \cite{HadronSpectrum:2012gic} and in bottomonium \cite{Ryan:2020iog} and the same adjustment of $c_s$ may also improve the relative positions of the spin-singlet and spin-triplet states with $J^P = (0,1,2)^+$~\cite{HadronSpectrum:2012gic}. Other splittings determined in this study are in reasonable agreement with their experimentally measured values.

\begin{table}
\begin{minipage}[t]{0.48\linewidth}
\begin{tabular}[t]{c| l l}
$B$ & this work & PDG \\ \hline
$B^{*} - B$  & $39 \pm 2 $ & $45.21 \pm 0.21$ \\
$B_1 - B^{*}$ & $413 \pm 6$ & $401.4 \pm 1.2$ \\
$B_1 - B$ & $452 \pm 6$ & $446.7 \pm 1.3$ \\
$B_2^{*} - B$ & $494 \pm 6$ & $457.5 \pm 0.7$ \\
$B_2^{*} - B_1$ & $42 \pm 8$ & $13.4 \pm 1.4$ \\
\end{tabular}
\end{minipage}%
\begin{minipage}[t]{0.48\linewidth}

\begin{tabular}[t]{c| l l}
$B_s$ & this work  & PDG  \\ \hline
\hspace{.1825cm} $B_s^{*} - B_s$ \hspace{.1825cm} & $38 \pm 1$ & $48.5^{+1.8}_{-1.5}$ \\
$B_{s2}^{*} - B$ & $554 \pm 4$ & $560.52 \pm 0.14$ \\
\end{tabular}

\end{minipage}

\centering
\begin{tabular}{c| l l}
$B_c$ & this work  & PDG  \\ \hline
$B_c^{*} - B_c$ & $\it 33.7 \pm 0.3$ & \\
$B_c (2s) - B_c $ & $602 \pm 3$ & $597.73 \pm 1.27$ \\
$B_c - B_s$ & $906 \pm 1$ & $907.75 \pm 0.37 \pm 0.27$ \\
\end{tabular}
\caption{A selection of mass splittings in MeV for $B$ (top left), $B_s$ (top right) and $B_c$ (bottom).
  The quoted uncertainties are statistical only and the italicised value for $B_c^{*} - B_c$ highlights that the very small error is likely an underestimate of the true uncertainty. PDG values are from the Particle Data Group \cite{ParticleDataGroup:2022pth}.}
\label{Tab:Splittings} 
\end{table}

\subsection{Hybrid Supermultiplets}
\label{sec:hybrid_supermultiplets}
In earlier calculations using the same lattices but with other quark flavours~\cite{Dudek:2009qf,Dudek:2010wm,Dudek:2011tt,Edwards:2011jj,Dudek:2011bn,Dudek:2012ag,HadronSpectrum:2012gic,Moir:2013ub,Padmanath:2013zfa,Padmanath:2015jea,Cheung:2016bym,Ryan:2020iog}, evidence for hybrid-like states has been consistently observed with a lightest hybrid state lying approximately 1.2 GeV above the lightest pseudoscalar for all flavours, from light to bottom. Where $C$-parity is present, a hybrid supermultiplet of nearly-degenerate states with $J^{PC}=(0,1,2)^{-+},1^{--}$ is also observed consistently. Since $C$-parity is broken for heavy-light mesons there are no quark-model exotic quantum numbers, nonetheless similar patterns are observed.

By examining the operator-state overlaps for each energy level, candidate states for a lightest hybrid supermultiplet for each heavy-light flavour were identified. The candidate states are shown in lighter colours in Figures~\ref{fig:IrrepSpectra} and~\ref{fig:IrrepSpectraBc} and in red in
Figures~\ref{fig:ContSpectra} and~\ref{fig:ContSpectra2}.

\begin{figure}[tb]
	\centering
\includegraphics[width=0.99\textwidth]{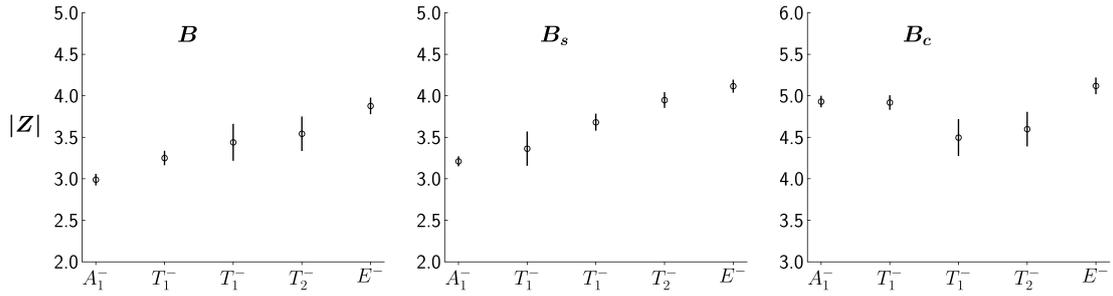}
\caption{Overlaps of the simplest operator proportional to the field-strength tensor, $ \left( (\pi , \rho) \times D^{[2]}_{J=1} \right)^J $,  for candidate hybrid states in $B$, $B_s$, and $B_c$. Since these are a similar magnitude across each irrep, and each corresponding energy level is of a similar mass as seen from the red boxes in Figure \ref{fig:HybridMass}, these are identified as a hybrid supermultiplet.}
\label{fig:Overlaps}
\end{figure}

\begin{figure}[tb]
\centering
\includegraphics[width=0.99\textwidth]{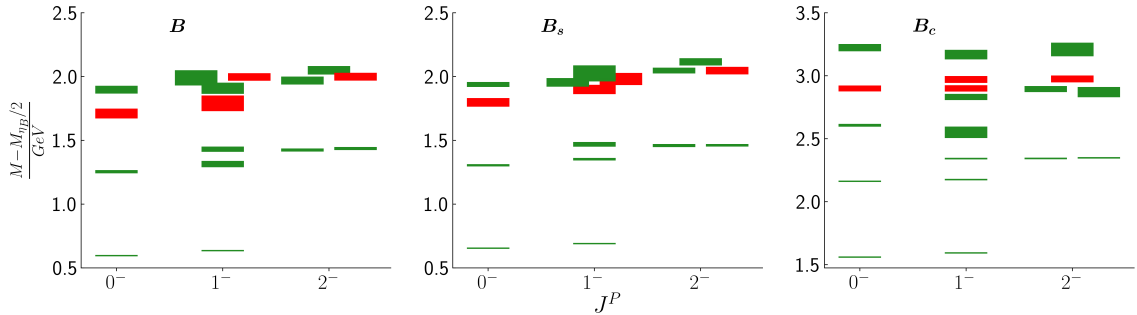}
\caption{Plots showing the continuum spins ($0^-$, $1^-$,$2^-$) where hybrid candidates were identified, in each of $B$, $B_s$ and $B_c$. Hybrid candidate states are shown in red.}
\label{fig:HybridMass}
\end{figure}

A state is proposed to be a hybrid meson if it has a relatively large overlap onto an operator proportional to the field-strength tensor. In Figure~\ref{fig:Overlaps} the overlap of the simplest hybrid operator construction $\left( (\pi , \rho) \times D^{[2]}_{J=1} \right)^J$ on each of the candidate states is shown, following the analysis in Ref.~\cite{Dudek:2010wm}. For each meson, the hybrid-like operator overlap is consistent across irreps (continuum spins). As a result, for each meson flavour, a supermultiplet of hybrid states with continuum quantum numbers $J^P = (0,1,2)^-$ is identified. The energy levels (hybrids) corresponding to these continuum spins are plotted separately in Figure~\ref{fig:HybridMass} for each of $B$, $B_s$ and $B_c$.

\subsection{Mixing of spin-singlet and spin-triplet states}
Since heavy-light mesons are not eigenstates of charge conjugation, spin-triplet and spin-singlet states with $J = L$ can mix
(in particular, states with $^3 L_{J=L} $ and $^1 L_{J=L}$). One can attempt to quantify this using mixing angles.  Following section~6.2 of Ref.~\cite{Moir:2013ub}, we consider a two-state hypothesis, with the assumption that energy independent mixing can occur. States $X$ and $Y$ may then be expanded in terms of spin-singlet and spin-triplet basis states,  
 \begin{equation}
\begin{aligned}
& |X \rangle & = +  \cos \theta \; |^1 L_{J=L} \rangle & + \sin \theta \; |^3 L_{J=L} \rangle , \\
& |Y \rangle & = -  \sin \theta \; |^1 L_{J=L} \rangle & + \cos \theta \; |^3 L_{J=L} \rangle ,
\end{aligned}
\end{equation}
with the choice $m_Y$ $>$ $m_X$. In the non-relativistic limit, there are certain operators that only overlap onto either spin-triplet or spin-singlet states. These contain the expression $\left[ (\rho - \rho_2) \right]$ for spin-triplet operators, and one of either $\left[ \pi \right]$ or $\left[ \pi_2 \right]$ for spin-singlet operators. By taking a ratio of the overlap of these operators on states $X$ and $Y$, one can determine the mixing angle $\theta$ (specifically $\tan \theta$ or $\cot \theta$). This angle gives an estimate of how mixed the spin-singlet and spin-triplet states are on the lattice, with $\theta = 45^{\circ}$ corresponding to maximal mixing, while $\theta = 0^{\circ},90^{\circ}$ corresponds to zero mixing.

Table~\ref{tab:HLMix} lists the absolute values of the mixing angles determined between the lightest pairs of $P$-wave ($1^+$), $D$-wave ($2^-$) and hybrid ($1^-$) states in $B$ and $B_s$. Also shown in this table are the mixing angles for the $D$ and $D_s$ mesons, determined in a previous study~\cite{Moir:2013ub}. The heavy-quark limit for the mixing of the lightest pair of $P$-wave and $D$-wave states~\cite{Close:2005se} is also shown.
The mixing angles show some mild dependence on the choice of operator which may be related to the underlying assumptions about singlet-triplet mixing or to the presence of lattice artefacts.

\begin{table}
\centering
\begin{tabular}{c|c|c c c c }
& & \multicolumn{3}{c}{$| \theta | / ^{\circ}$} \\
& $Q$-$q$ & $ \sim (\rho - \rho_2 )$ & $ \sim \pi$ & $ \sim \pi_2$ \\ \hline\hline
$1^+$ & $c$-$\bar{s}$ & 60.9(0.6) & 64.9(0.2) & 66.4(0.4) & ~\cite{Moir:2013ub}\\
 & $c$-$\bar{\ell}$ & 60.1(0.4) & 62.6(0.2) & 65.4(0.2) & ~\cite{Moir:2013ub}\\
 & $\bar{b}$-$s$ & 59.3(0.6) & 62.9(0.4) & 63.5(0.4) \\
 & $\bar{b}$-$\ell$ & 58.7(0.6) & 61.1(0.6) & 62.0(0.6) \\  \hline
 & H.Q.L. & \multicolumn{3}{c}{54.7 or 35.3} \\ \hline\hline
$2^-$ & $c$-$\bar{s}$ & 64.9(1.9) & 68.7(2.0) & 70.9(1.8) & ~\cite{Moir:2013ub}\\
 & $c$-$\bar{\ell}$ & 63.3(2.2)$^{*}$ & 67.8(3.7)$^{*}$ & 71.1(3.9)$^{*}$ & ~\cite{Moir:2013ub}\\
 & $\bar{b}$-$s$ & 55.6(1.1) & 61.5(0.9) & 62.2(0.9) \\
 & $\bar{b}$-$\ell$ & 52.4(0.5) & 57.7(0.5) & 58.7(0.5) \\  \hline
 & H.Q.L. & \multicolumn{3}{c}{50.8 or 39.2} \\ \hline\hline
$1^-$ (hybrid) & $c$-$\bar{s}$ & 59.9(1.7) & 67.9(0.9) & 67.3(0.9) & ~\cite{Moir:2013ub}\\
 & $c$-$\bar{\ell}$ & 59.7(1.1) & 68.4(0.8) & 67.4(0.9) & ~\cite{Moir:2013ub}\\
 & $\bar{b}$-$s$ & 60.3(1.1) & 61.6(1.2) & 61.3(1.2) \\
 & $\bar{b}$-$\ell$ & 57.3(1.4) & 65.1(1.1) & 64.7(1.2) \\  \hline
 & H.Q.L. & \multicolumn{3}{c}{$-$} \\ \hline\hline
\end{tabular}
\caption{The absolute value of the mixing angles calculated for the lightest pairs of $P$-wave ($1^+$), $D$-wave ($2^-$) and hybrid ($1^-$) states of the $\bar{b}$-$\ell$ and $\bar{b}$-$s$ mesons, as well as the $c$-$\bar{\ell}$ and $c$-$\bar{s}$ results presented in Ref.~\cite{Moir:2013ub}. Angles extracted using different operators are presented. H.Q.L. denotes results determined in the heavy-quark limit~\cite{Close:2005se}. The results marked with $^{*}$ correspond to $(90-| \theta |)$, to account for a difference with the
  mass ordering in the open-charm meson results~\cite{Moir:2013ub}.}
\label{tab:HLMix}
\end{table}

\begin{figure}[t]
\centering
	\begin{subfigure}{0.48\textwidth}
		\includegraphics[width=\textwidth]{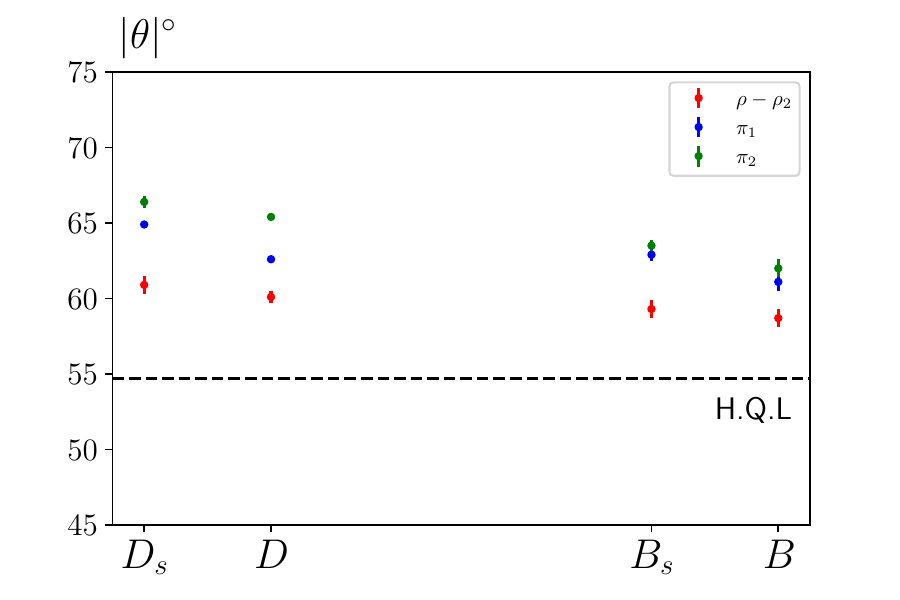}
		\subcaption{$P$-wave}
	\end{subfigure}
	\begin{subfigure}{0.48\textwidth}
		\includegraphics[width=\textwidth]{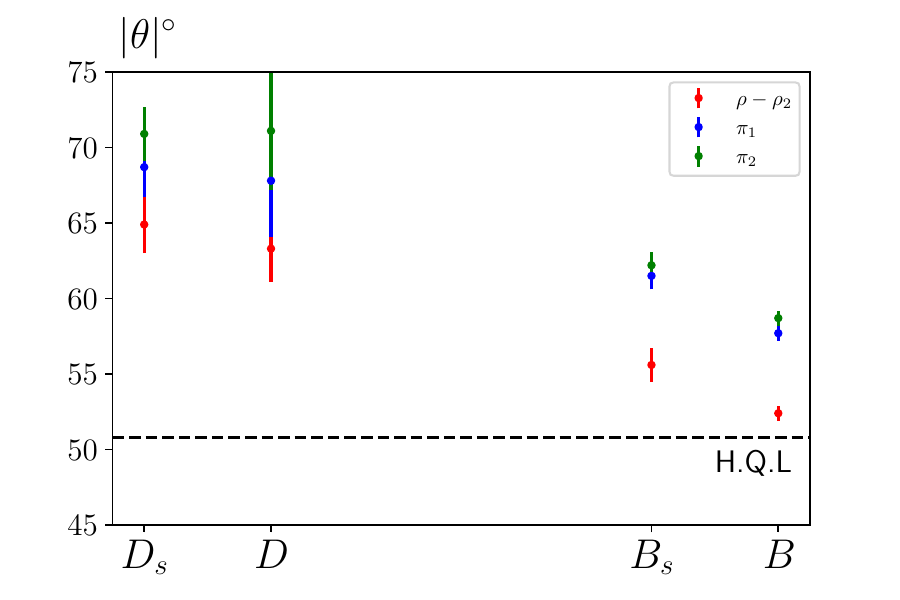}
		\subcaption{$D$-wave}
	\end{subfigure}
\caption{The absolute value of the mixing angles calculated for the lightest (a) $P$-wave and (b) $D$-wave spin-singlet and spin-triplet state pairs ordered by relative quark mass difference. Each choice of operator has been coloured differently with red for $\sim(\rho - \rho_2 )$, blue for $\sim \pi$ and green for $\sim \pi_2$. The mixing angles in the heavy-quark limit are shown as dashed lines.}
\label{fig:mix}
\end{figure}

The mixing angles determined for the $\bar{b}$-$\ell$ and $\bar{b}$-$s$ mesons are found to be similar. This may be expected given the heavier-than-physical light quarks and close-to-physical strange quarks that were used, meaning that SU(3) flavour symmetry is not badly broken. Therefore, due to the unphysical light quark mass, one might expect the $\bar{b}$-$s$ meson results to be closer to their physical value than the $\bar{b}$-$\ell$.
Turning to the heavy quark dependence, the values of $|\theta|$ for the $P$-wave and $D$-wave states lie between predictions in the heavy-quark limit and the mixing angles determined for charmed $c$-$\bar{\ell}$ and $c$-$\bar{s}$ mesons. The absolute values of the mixing angles for the lightest $P$-wave and $D$-wave spin-singlet and spin-triplet state pairs, ordered by relative quark mass difference, are shown in Figure~\ref{fig:mix}. The plot illustrates the trend towards the heavy-quark limit and also highlights the systematic uncertainty introduced by the choice of operator.

Table \ref{tab:BcMix} shows the absolute value of the mixing angles determined for the lightest pairs of $P$-wave ($1^+$), $D$-wave ($2^-$) and hybrid ($1^-$) states in $B_c$. In this case, the mixing angles are significantly larger than those determined for the heavy-(light,strange) mesons and further from the heavy-quark limit. It seems that in this case, the separation of scales between the bottom and charm quarks may not be large enough for the heavy-quark limit to apply. Instead, bottom-charm mesons appear to be approaching the results of quarkonium, where zero mixing occurs and the spin-singlet and spin-triplet states can be differentiated. The mixing angles determined here favour this hypothesis, being closer to zero mixing than any of the other mesons, especially the mixing of the $D$-wave ($J^P = 2^-$) spin-singlet and spin-triplet states.

\begin{table}[t]
\centering
\begin{tabular}{c|c|c c c}
\hline
& & \multicolumn{3}{c}{$| \theta | / ^{\circ}$} \\
& $J^P$ &  $ \sim (\rho - \rho_2 )$ & $ \sim \pi$ & $ \sim \pi_2$ \\ \hline
$\bar{b}$-$c$ & $1^+$ & 78.6(0.5) & 79.1(0.5) & 79.1(0.5) \\
 & $2^-$ & 88.0(0.9) & 89.4(0.4) & 89.5(0.4) \\
 & $1^{-}$ (hybrid) & 69.4(1.4) & 71.3(1.2) & 71.2(1.2) \\ \hline
\end{tabular}
\caption{The absolute value of the mixing angles calculated for the lightest pairs of the $P$-wave ($1^+$), $D$-wave ($2^-$) and hybrid ($1^-$) spin-singlet and spin-triplet $B_c$ states. Angles extracted using different operators are presented.}
\label{tab:BcMix}
\end{table}

\begin{figure}[t]
\centering
	\begin{subfigure}{0.8\textwidth}
		\includegraphics[width=\textwidth]{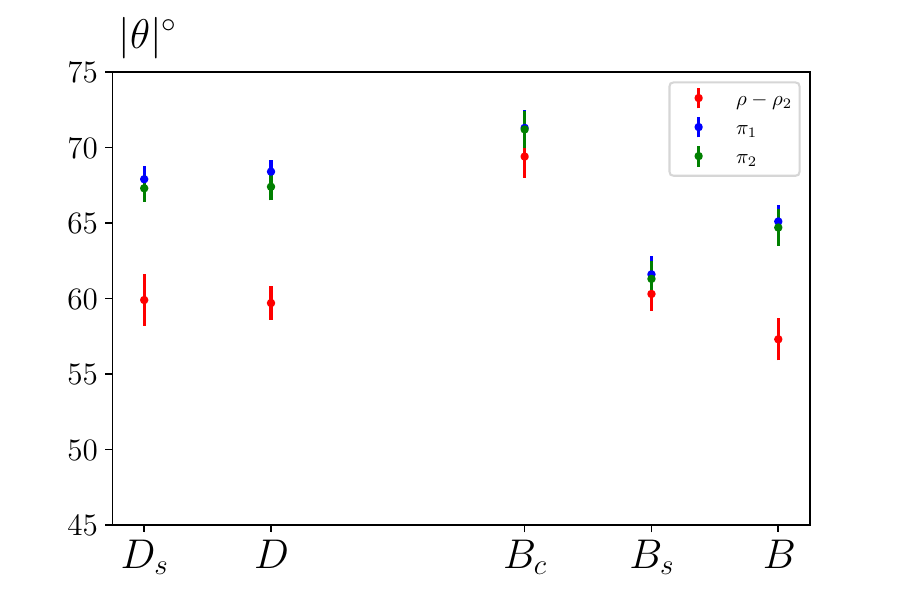}
	\end{subfigure}
	\caption{The absolute value of the mixing angles calculated for the lightest hybrid ($1^-$) spin-singlet and spin-triplet state pairs ordered by relative quark mass difference. Each choice of operator has been coloured differently with red for $\sim (\rho - \rho_2 )$, blue for $\sim \pi$ and green for $\sim \pi_2$.}
	\label{fig:mixHybrid}
\end{figure}

Turning to the hybrid mesons, a plot of the absolute value of the mixing angles for the lightest
hybrid ($1^-$) spin-singlet and spin-triplet state pairs, for each flavour sector, ordered by relative quark mass difference, is shown in Figure \ref{fig:mixHybrid}. This plot suggests that the mixing angle for the hybrid spin-singlet and spin-triplet states does not depend as strongly on constituent quark masses as it does
for conventional heavy-light mesons.

In general, the systematic uncertainties from the choice of operators are quite significant for some of the mesons, while mixing angles for the bottom-charm and bottom-strange depend less on the choice of operator used to determine the mixing angle.

\subsection{Comparisons to other studies}
The heavy-light meson spectrum has been studied previously in quark models and in lattice calculations with various approaches to the heavy quark mass. Here we discuss only selected comparisons and highlight some interesting features of the spectra determined in this study.

The utility of the operator overlaps in state identification has already
been illustrated in Section~\ref{sec:hybrid_supermultiplets}. A similar analysis of the overlaps also
allows for the identification of multiplets following the notation of quark
models. The method is especially useful for closely clustered energy levels
in the same $J^P$ channel. As an example, note that the first and second
excited states in $J^P=2^+$ in each of $B, B_s$ and $B_c$ lie close together at
$\sim 1.75$ GeV for $B$ and $B_s$ and $\sim 2.65$ GeV for $B_c$, as shown in Figures~\ref{fig:ContSpectra}
and~\ref{fig:ContSpectra2}. Additional information is needed to determine which energy level is a member of the $P$-wave $(0^+,1^+,2^+)$ triplet and
which is in the $F$-wave $(2^+,3^+,4^+)$ triplet of states. By comparing the operator
overlaps to the other $J^P$ states in the $P$-wave and $F$-wave multiplets we find that
in each case the lower of the two energy levels can be assigned to the $P$-wave triplet
while the higher-lying state is part of the $F$-wave. The same analysis allows us to identify the first excited state in $J^P=1^-$ as $S$-wave and the nearby second excited state as a part of a $D$-wave $(1^-,2^-,3^-)$ multiplet, in each of $B, B_s$ and $B_c$.

Comparing to early quark model predictions of $B$ and $B_s$ states~\cite{Godfrey:1985xj}, a similar pattern of states and level orderings are observed in this work. Later quark model predictions in the
$B_c$ spectrum~\cite{Godfrey:2004ya} are also broadly compatible.
We find generally good agreement between the ground states determined in this work and those
in quark models. However, above strong decay thresholds some differences emerge. In particular,
the higher-lying energy levels determined in this work tend to lie above the same states
in quark model calculations in $B$, $B_s$ and $B_c$ in Refs.~\cite{Lahde:1999ih, Ebert:2009ua, DiPierro:2001dwf, Sun:2014wea, Lu:2016bbk, Ortega:2016pgg}. However, care must be taken in these comparisons where the
effects of thresholds and the position of energy levels with respect to those thresholds
should first be understood.
This study has identified hybrid mesons, and in particular a hybrid supermultiplet of states at a common energy above the ground state, in each of $B, B_s$ and $B_c$ and which does not feature in quark model predictions.

Recent predictions of $B_c$ states~\cite{Martin-Gonzalez:2022qwd, Ortega:2020uvc} using
a constituent non-relativistic quark model yielded a similar pattern of states to the 
spectrum determined in this work. In this quark model calculation two $S$-wave multiplets,
two $P$-wave multiplets, and a $D$-wave multiplet are predicted to lie below the
$BD$ threshold. The same multiplets are determined here with the second $P$-wave multiplet, consisting of states with $J^P$ = $(0^+,1^+,2^+)$, found above the $BD$ threshold. We emphasise again that a full understanding of threshold effects can only be elucidated by the inclusion of (appropriate) meson-meson operators. 

The low-lying positive parity states in $B$ and $B_s$ may be of particular interest as they
are close to relevant strong decay thresholds.
A recent calculation of the $B_{s0}^*$ and $B_{s1}$ ground states
using lattice NRQCD is described in Ref.~\cite{Hudspith:2023loy}. The results presented here
are in good agreement with the same states determined in that study. In this work, we find
the masses of the $B_{s0}^*$ and $B_{s1}$ are 79 MeV and 103 MeV below the lattice $BK$ and
$B^*K$ thresholds, respectively. In Ref.~\cite{Hudspith:2023loy} similar energy levels
are found with respect to thresholds with a caveat that systematic uncertainties are dominated
by discretization effects.

In this study, some systematic uncertainties remain unquantified. Nevertheless, the results
are consistent with previous results and with experimental results, while also providing the first  
extensive study of highly-excited and hybrid heavy-light mesons. It also hints at the $J^P$ channels and
energy regions that might be interesting for further study.

\section{Summary}
\label{sec:sum}

We have presented excited-state spectra of $B$, $B_s$ and $B_c$ mesons computed from lattice QCD. These have been determined with $N_f=2+1$ flavours of dynamical quarks on an anisotropic lattice at a single spatial lattice spacing, and with a light quark mass corresponding to $m_\pi\approx 391$~MeV. Using a large basis of operators, an extensive spectrum of states is identified up to $J=4$. For the lower-lying excited states we find patterns similar to results from quark potential models. At higher energies, approximately 1500 MeV above the ground state in each flavour sector, a hybrid supermultiplet is identified comprising states with $J^P = (0,1,2)^-$.

Previous calculations in the same approach by the Hadron Spectrum Collaboration have considered all other flavour combinations in mesons
with light, strange~\cite{Dudek:2009qf,Dudek:2010wm,Dudek:2011tt,Edwards:2011jj,Dudek:2011bn,Dudek:2012ag}, and charm~\cite{HadronSpectrum:2012gic,Moir:2013ub,Cheung:2016bym} quarks, while the spectrum of $b\bar{b}$ mesons was described in Ref.~\cite{Ryan:2020iog}.
In each case, a qualitatively similar pattern of states including patterns resembling hybrid-like states was found.

A full error budget was beyond the scope of this work. Certainly, in simulations of $b$-quarks discretization effects are of concern and the impact of these effects was explored indirectly by comparing fits to the continuum relativistic dispersion relation between the heavy-light mesons and to dispersion relations obtained previously in bottomonium. The measured anisotropy -
related to the speed of light - was consistent across fits, including to quite high momenta. The anistropies were also consistent with
the value determined from the pion dispersion relation, while a mild tension was noted between $b$-$\bar{b}$ and $\bar{b}$-$\ell$.

Compared with the charmed $D$ and $D_s$ mesons, the $B$ and $B_s$ mesons are relatively unexplored but it appears that many
as-yet unobserved states may be awaiting experiments. This work signals that there are interesting hybrid-like states high in the
spectrum of $B, B_s$ and $B_c$ mesons, and also suggests that near-threshold scalar states may arise, again in each of
$B$, $B_s$ and $B_c$. The latter are an interesting prospect for further lattice studies while comparisons to
$D$-mesons will help in understanding the interactions of QCD at work at these energies.

\bigskip

\begin{acknowledgments}
We thank our colleagues within the Hadron Spectrum Collaboration (\url{www.hadspec.org}).
LG acknowledges funding from an Irish Research Council Government
of Ireland Postgraduate Scholarship [GOIP/2019/4446].
SMR acknowledges support from a Science Foundation Ireland Frontiers for the Future Project award [grant number SFI-21/FFP-P/10186]
DJW acknowledges support from a Royal Society University Research Fellowship, and support from the U.K. Science and Technology Facilities Council (STFC) [grant numbers ST/T000694/1 \& ST/X000664/1].
The software codes
{\tt Chroma}~\cite{Edwards:2004sx}, {\tt QUDA}~\cite{Clark:2009wm,Babich:2010mu}, {\tt QUDA-MG}~\cite{Clark:SC2016}, {\tt QPhiX}~\cite{ISC13Phi}, {\tt MG\_PROTO}~\cite{MGProtoDownload}, {\tt QOPQDP}~\cite{Osborn:2010mb,Babich:2010qb}, and {\tt Redstar}~\cite{Chen:2023zyy} were used. 
Some software codes used in this project were developed with support from the U.S.\ Department of Energy, Office of Science, Office of Advanced Scientific Computing Research and Office of Nuclear Physics, Scientific Discovery through Advanced Computing (SciDAC) program; also acknowledged is support from the Exascale Computing Project (17-SC-20-SC), a collaborative effort of the U.S.\ Department of Energy Office of Science and the National Nuclear Security Administration.

This work used the Cambridge Service for Data Driven Discovery (CSD3), part of which is operated by the University of Cambridge Research Computing Service (www.csd3.cam.ac.uk) on behalf of the STFC DiRAC HPC Facility (www.dirac.ac.uk). The DiRAC component of CSD3 was funded by BEIS capital funding via STFC capital grants ST/P002307/1 and ST/R002452/1 and STFC operations grant ST/R00689X/1. Other components were provided by Dell EMC and Intel using Tier-2 funding from the Engineering and Physical Sciences Research Council (capital grant EP/P020259/1). This work also used the earlier DiRAC Data Analytic system at the University of Cambridge. This equipment was funded by BIS National E-infrastructure capital grant (ST/K001590/1), STFC capital grants ST/H008861/1 and ST/H00887X/1, and STFC DiRAC Operations grant ST/K00333X/1. DiRAC is part of the National E-Infrastructure.
This work also used clusters at Jefferson Laboratory under the USQCD Initiative and the LQCD ARRA project.

Propagators and gauge configurations used in this project were generated using DiRAC facilities, at Jefferson Lab, and on the Wilkes GPU cluster at the University of Cambridge High Performance Computing Service, provided by Dell Inc., NVIDIA and Mellanox, and part funded by STFC with industrial sponsorship from Rolls Royce and Mitsubishi Heavy Industries. Also used was an award of computer time provided by the U.S.\ Department of Energy INCITE program and supported in part under an ALCC award, and resources at: the Oak Ridge Leadership Computing Facility, which is a DOE Office of Science User Facility supported under Contract DE-AC05-00OR22725; the National Energy Research Scientific Computing Center (NERSC), a U.S.\ Department of Energy Office of Science User Facility located at Lawrence Berkeley National Laboratory, operated under Contract No. DE-AC02-05CH11231; the Texas Advanced Computing Center (TACC) at The University of Texas at Austin; the Extreme Science and Engineering Discovery Environment (XSEDE), which is supported by National Science Foundation Grant No. ACI-1548562; and part of the Blue Waters sustained-petascale computing project, which is supported by the National Science Foundation (awards OCI-0725070 and ACI-1238993) and the state of Illinois. Blue Waters is a joint effort of the University of Illinois at Urbana-Champaign and its National Center for Supercomputing Applications.
\end{acknowledgments}

\section*{Data access}

Reasonable requests for data can be directed to the authors and will be considered in accordance with the Hadron Spectrum Collaboration's policy on sharing data.

\bibliography{B_Bs_Bc_spectrum_paper.bib}
\bibliographystyle{JHEP}

\end{document}